\documentstyle[12pt,epsfig,rotating]{article}
\def\figcap{\section*{Figure Captions\markboth
        {FIGURECAPTIONS}{FIGURECAPTIONS}}\list
        {Figure \arabic{enumi}:\hfill}{\settowidth\labelwidth{Figure
999:}
        \leftmargin\labelwidth
        \advance\leftmargin\labelsep\usecounter{enumi}}}
 \relax
\def\ap#1#2#3{Ann.\ Phys.\ (NY) #1 (19#3) #2}

\def\np#1#2#3{Nucl.\ Phys.\ B#1 (19#3) #2}
\def\pl#1#2#3{Phys.\ Lett.\ #1B (19#3) #2}

\def\prb#1#2#3{Phys.\ Rev.\ B #1 (19#3) #2}
\def\prep#1#2#3{Phys.\ Rep.\ #1 (19#3) #2}

\def\rmp#1#2#3{Rev.\ Mod.\ Phys.\ #1 (19#3) #2}
\newcounter{hran}
\def\bmini#1{\setcounter{hran}{\value{equation}}
\refstepcounter{hran} \setcounter{equation}{#1}
\renewcommand{\theequation}{\thehran\alph{equation}}
              \begin{eqnarray}  }
\def\bminiG#1{
          \setcounter{hran}{\value{equation}}
          \refstepcounter{hran}
          \setcounter{equation}{-1}
          \renewcommand{\theequation}{\thehran\alph{equation}}
          \refstepcounter{equation}
    \label{#1}
          \begin{eqnarray}          }
\def\emini{\end{eqnarray}\setcounter{equation}{\value{hran}}
\renewcommand{\theequation}{\arabic{equation}}}
\newskip\humongous \humongous=0pt plus 1000pt minus 1000pt
\def\caja{\mathsurround=0pt} \def\eqalign#1{\,\vcenter{\openup1\jot
\caja   \ialign{\strut \hfil$\displaystyle{##}$&$
\displaystyle{{}##}$\hfil\crcr#1\crcr}}\,} \newif\ifdtup

\def\re#1{(\ref{#1})}
\def\half{\mbox{\small $\frac{1}{2}$}}

\def\frac#1#2{ {{#1} \over {#2} }}
\def\ie{\hbox{\it i.e.}{ }}      
      
\def\beq{\begin{equation}}
\def\eeq{\end{equation}}
\def\beeq{\begin{eqnarray}}
\def\eeeq{\end{eqnarray}}
\def\as{\alpha_s}
\def\bc{\bar c}

\def\G{\Gamma}
\def\bG{ \bar \Gamma}
\def\Gr{\G_{\mbox{\footnotesize{rel}}}}
\def\L{ \Lambda}

\def\g{ \gamma}
\def\tg{ \hat g}
\def\r{\rho}
\def\dr{\dot \r}
\def\dZ{\dot Z}
\def\br{\bar\r}
\def\rUV{\r^{\mbox{\scriptsize UV}}}

\def\brUV{\bar\rUV}
\def\U{{\mbox{\scriptsize UV}}}

\def\dL{\L \partial_\L }
\def\mRG{$\mu$-RG }
\def\lRG{$\L$-RG }
\def\bp{ \bar p}
\def\UV{$\L_0\to\infty\;$}
\def\bit{\begin{itemize}}
\def\eit{\end{itemize}}
\def\ben{\begin{enumerate}}
\def\een{\end{enumerate}}

\begin{document}
\begin{titlepage}
\renewcommand{\thefootnote}{\fnsymbol{footnote}}
\begin{flushright}
     UPRF 96-464\\
     IFUM-525-FT\\
     March 1996 \\
\end{flushright}
\par \vskip 10mm
\begin{center}
{\Large \bf
Beta function and infrared renormalons\\ 
in the exact Wilson renormalization group in Yang-Mills theory
\footnote{Research supported in part by MURST, Italy
and by EC Programme ``Human Capital and Mobility", 
contract CHRX-CT93-0357 (DG 12 COMA).}}
\end{center}
\par \vskip 2mm
\begin{center}
M.\ Bonini$\,^a$, 
G.\ Marchesini$\,^b$
and  M.\ Simionato$\,^a$ \\
\vskip 5 mm
$^a\,${\it Dipartimento di Fisica, Universit\`a di Parma \\
and INFN, Gruppo Collegato di Parma, Italy}\\
\vskip 2 mm
$^b\,${\it Dipartimento di Fisica, Universit\`a di Milano \\
and INFN, Sezione di Milano, Italy}
\end{center}

\par \vskip 2mm
\begin{center} {\large \bf Abstract} \end{center}
\begin{quote}
We discuss the relation between the
Gell-Mann-Low beta function and the ``flowing couplings'' of
the Wilsonian action $S_\L[\phi]$ of the exact renormalization
group (RG) at the scale $\L$.
This relation involves the ultraviolet region of $\L$ so that
the condition of renormalizability is equivalent to the 
Callan-Symanzik equation.
As an illustration, by using the exact RG formulation, we compute 
the beta function in Yang-Mills theory to one loop (and to two loops 
for the scalar case). We also study the infrared (IR) renormalons. 
This formulation is particularly suited for this study since: 
$i$) $\L$ plays the r\^ole of a IR cutoff in Feynman diagrams and
non-perturbative effects could be generated as soon as $\L$ becomes small;
$ii$) by a systematical resummation of higher order corrections
the Wilsonian flowing couplings enter directly into the Feynman 
diagrams with a scale given by the internal loop momenta;
$iii$) these couplings tend to the running coupling at high 
frequency, they differ at low frequency and remain finite all 
the way down to zero frequency.
\end{quote}
\end{titlepage}

\section{Introduction}
In renormalization group (RG) there are two different ways to
introduce a momentum scale.
The first leads to the Gell-Mann-Low beta function and Callan-Symanzik
equation, the second to the exact Wilson RG flow.
In the first formulation one introduces the scale 
$\mu$ at which the coupling $g$, mass and wave function normalization are
measured or fixed. The fact that physical quantities do not depend
on the scale $\mu$ leads to the Gell-Mann-Low beta function and the
Callan-Symanzik equation.
In this formulation one does not usually attribute any meaning to
the bare action.
In the Wilson RG formulation \cite{W} instead the starting point is
the local bare action $S_{\L_0}[\phi]$ in which the fields $\phi$ have 
frequencies smaller than the ultraviolet (UV) scale $\L_0$.
The UV action can be viewed as the result of taking a more elementary
theory and integrating all fields with frequencies larger than $\L_0$.
This gives the coefficients of $S_{\L_0}[\phi]$ such as the wave function 
normalizations $Z_i(\L_0)$ and the parameters $\r_i(\L_0)$.
By further integrating the fields with frequence larger than $\L<\L_0$ 
one obtains a new action $S_\L[\phi]$ which is not any more local.
This Wilsonian action, and in particular the coefficients $\r_i(\L)$
and $Z_i(\L)$ of its local part, the ``relevant parameters'', depends 
on $\L$ according to the exact RG equation \cite{W}-\cite{B}.
The physical observables are obtained by performing the path integration
over all fields with momenta in the full range $0<q^2<\L^2_0$.
Therefore for $\L=0$ (and $\L_0\to\infty$) the Wilsonian action 
generates the physical Green functions. In particular for the wave
function normalization one has $Z_i(0)=1$ and the parameters $\r_i(0)$
are given by the masses and physical coupling $g$ (at the scale $\mu$). 

In this paper we establish the relation between
the two RG formulations. We analyze, as examples, the cases of the
massless scalar and the $SU(2)$ Yang-Mills theories.
To stress the difference on the way the scale is introduced, hereafter
we shall call \mRG and \lRG the first and the second formulation.

First we deduce the Gell-Mann-Low beta function in terms of the
derivative of the relevant parameters $\r_i(\L)$ and $Z_i(\L)$ of the
Wilsonian action. Since RG is related to the UV properties, as expected,
this relation involves the UV region of $\L$. 
Although all perturbative contributions to the relevant parameters
diverge logarithmically for large $\L$, in the combination
giving the beta function all divergences cancel\footnote{
In Ref.~\cite{H} the beta functional has been obtained to one-loop order
in terms of the relevant parameters of the \lRG formulation.}.
It is interesting to observe that in the Yang-Mills case the UV
action in general does not satisfy the BRS symmetry and then it could 
be surprising that the beta function is given in terms of the 
UV parameters.
However one should observe that these UV parameters depend on $g$ and 
$\mu/\L_0$ in such a way that the physical effective action 
($\L=0$ and $\L_0\to\infty$) satisfies Slavnov-Taylor identities
\cite{B,BDMYM}.

As a second point concerning the relation between \mRG and \lRG
formulations we show that the Callan-Symanzik equation is equivalent
to the condition that the theory is renormalizable, \ie physical
vertices do not depend on the UV scale as $\L_0\to\infty$.
This is a special case of the \lRG equation which is based on the fact that
the physical vertices are obtained from the Wilsonian action
$S_\L[\phi]$ independently of the value of $\L$.

As a third point we apply the \lRG formulation to the study of 
infrared (IR) renormalons \cite{renormalons} in non-Abelian gauge 
theory (and triviality in the scalar theory).
IR renormalons could be related to the fact that in 
a Feynman diagram with a virtual momentum $k$ there are higher order 
corrections which reconstruct the running coupling $g(k)$ at 
the scale $k$. 
Then the momentum integration involves small values of $k^2$ where 
the perturbative running coupling becomes ambiguous and one must 
include non-perturbative corrections to avoid the IR Landau pole.
Although in non-Abelian theories there are various indications 
supporting this higher order resummations leading to $g(k)$ 
(non-Abelianization \cite{na} and dispersive methods \cite{disp}), 
no systematic proof of this is known. 
The \lRG formulation provides a suitable framework in 
which one can reconstruct the running coupling at large loop 
momenta and one can study the IR region in Feynman diagrams.
The natural couplings in \lRG are the {\it Wilsonian flowing couplings} 
$\tg_i(\L)$ given by the relevant parameters $\r_i(\L)$ rescaled by 
the appropriate wave functions $Z_j(\L)$. 
In the Yang-Mills case there are three Wilsonian couplings $\tg_i(\L)$, 
the relevant coefficients of the three interacting monomials in the 
classical action (in the scalar case one has a single $\tg(\L)$). 

We show that it is possible to set up an iterative procedure to 
solve the \lRG equations in such a way that one generates 
contributions of Feynman diagrams in which the loop momenta 
$k_i$ are ordered in virtuality ($k^2_1<\cdots k^2_\ell$) 
and in which the integrands involve the flowing couplings 
$\tg_i(k_j)$ at these momentum scales. 
The flowing couplings themselves can be obtained by a generalized beta 
function which can be computed as an expansion in the Wilsonian 
coupling itself.
We explicitly compare the Wilsonian and the running couplings in the 
one-loop resummation. 

At large scales the running $g(k)$ and the Wilsonian $\tg_i(k)$ 
couplings differ only by $\mu/k$ power corrections.
Therefore increasing $k$ all couplings vanish in the Yang-Mills case,
while in the scalar case they grow and become singular at 
the UV Landau pole $k_L \simeq \mu\;exp( 1/b_0g)$ with $b_0$ the 
(positive) one-loop coefficient of the beta function. Therefore also 
within this \lRG analysis one finds the property of triviality.

For a $k$ comparable to $\mu$ the Wilsonian couplings differ from 
the running coupling.
In the non-Abelian case the one-loop running coupling $g(k)$ diverges 
at the IR Landau pole and the integration over the small $k$ region 
become meaningless.
In the \lRG formulation instead the Wilsonian couplings $\tg_i(k)$ 
remain finite in all the range of $k$ and at $k=0$ one has $\tg_i(0)=g$. 
Therefore the integration over $k$ is now possible. 

In the \lRG formulation $\L$ plays the role of an IR cutoff in Feynman 
diagrams. So, as long as $\L$ is much larger than $\mu$ one can use
the perturbative expressions for the various flowing couplings
$\tg_i(k_i)\simeq g(k_i)$.
The Landau singularity is cutoff by
the IR scale $\L$ so that non perturbative contributions are
negligible and of order $\mu/\L$.
For $\L$ approaching the physical point $\L=0$ one has that the
various flowing couplings $\tg_i(k_i)$ depart from the running 
coupling and power corrections resum to make the flowing couplings finite. 

The paper is organized as follows.
The details of our analysis are presented in Sect.~2 in the 
case of the massless scalar theory. In Sect.~3 we generalize this 
analysis to the Yang-Mills theory. Sect.~4 contains some conclusions.
In the Appendix we recall the form of the \lRG equations 
for the scalar case and describe how the usual loop expansion is 
obtained.

\section{Massless scalar case}

We illustrate in this case the main elements of the analysis. 
Some results are general and it is easier to illustrate them 
in the scalar case. 
We start by recalling the needed elements of \lRG and 
then we deduce in this formulation the beta function and compute the 
two-loop contribution. Then we introduce the improved perturbation 
expansion in terms of the flowing coupling and compare it with the 
usual running coupling. All these points will be reanalyzed in the 
Yang-Mills case in the next section. 

\subsection{Summary of \lRG formulation}

We recall the \lRG formulation by taking as example the euclidean
$\phi^4$ massless scalar theory in four dimension.
One starts from the action at an UV scale $\L_0$ larger than any physical
scale in the theory
\beq\label{S0}
S_{\L_0}[\phi]=
\int d^4 x
\left\{
\half
\phi(x)\;[-Z^\U \partial^2\;+\;\L_0^2\;\rUV_1]\;\phi(x)
+ \frac{\rUV_2}{4!}\; \phi^4(x)
\right\}
\,.
\eeq
All fields in the path integrals have frequencies smaller than
$\L_0$. Thus in Feynman diagrams all internal Euclidean momenta $q_i$
are bounded by $q_i^2 \le \L_0^2$.
The UV parameters $\rUV_i$ are dimensionless. All other possible
field monomials have coefficients which are suppressed by inverse
powers of $\L_0$ since they have negative dimension and here they are
neglected for simplicity.
Perturbatively this theory is renormalizable so that the UV parameters
can be made dependent on $\L_0$ in such a way that observables at the
physical scales are independent of $\L_0$ for $\L_0\to\infty$.

The dependence of the UV action on $\L_0$ can be generalized to smaller
values of the scale. Such a generalization, which is the basis for the
\lRG formulation \cite{W}, is obtained by performing the path
integrations over the fields with frequencies between $\L_0$ and $\L <\L_0$.
This gives the Wilsonian action $S_\L[\phi]$ in which
the coefficients of the various field monomials depend on $\L$.
Of course this action is not any more local.

An equivalent way to obtain the Wilsonian action follows from the
observation that $S_\L[\phi]$ generates Green functions
in which the propagators have both the UV and an IR cutoff $\L$
\beq\label{prop}
D_{\L\L_0}(p) \equiv \frac{K_{\L\L_0}(p) }{p^2}
\,,
\eeq
with $K_{\L\L_0}(p)=1$ in the region $\L^2\le p^2\le\L^2_0$
and rapidly vanishing outside.
From this property one deduces \cite{W,P} the RG evolution equation for
the Wilsonian action
\beq\label{EvEq}
\dL\; e^{ -S_\L[\phi]}\;=\; \half \int_q \;\dL D_{\L\L_0}(q)
\;\frac{\delta^2}{\delta \phi(q)\delta\phi(-q)}
\; e^{-S_\L[\phi]}
\,, \;\;\;\;\;\;\;\;
\int_q \equiv \int\frac{d^4q}{(2\pi)^4}
\,.
\eeq
For our study it is convenient to perform the Legendre transform of
the Wilsonian action $S_\L[\phi]$ and introduce the
``cutoff effective action'' $\G[\phi;\L,\L_0]$ generating the
``cutoff vertices'' $\G_{2n}(p_1,\cdots,p_{2n};\L,\L_0)$.
The RG equations for these vertices, corresponding to \re{EvEq}, are
schematically given in fig.~1 and they are recalled with the proper details 
in the Appendix (see \cite{BDM}-\cite{Mo}).
\begin{figure}
  \begin{center}
    \begin{tabular}{c}
    \epsfig{file=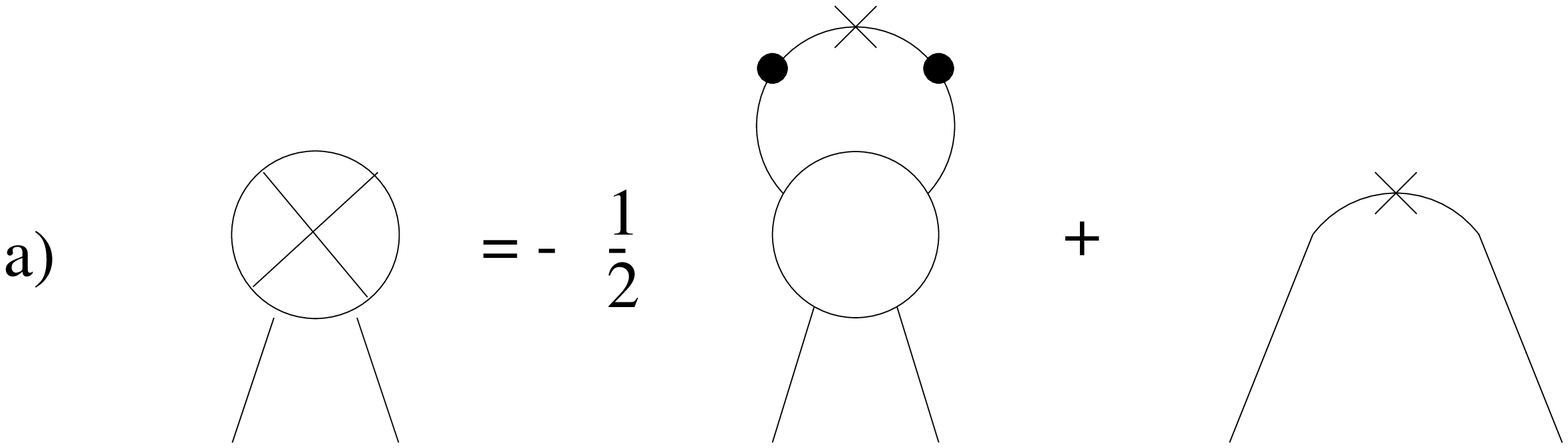,height=12 ex}\\
    {}\\
    \epsfig{file=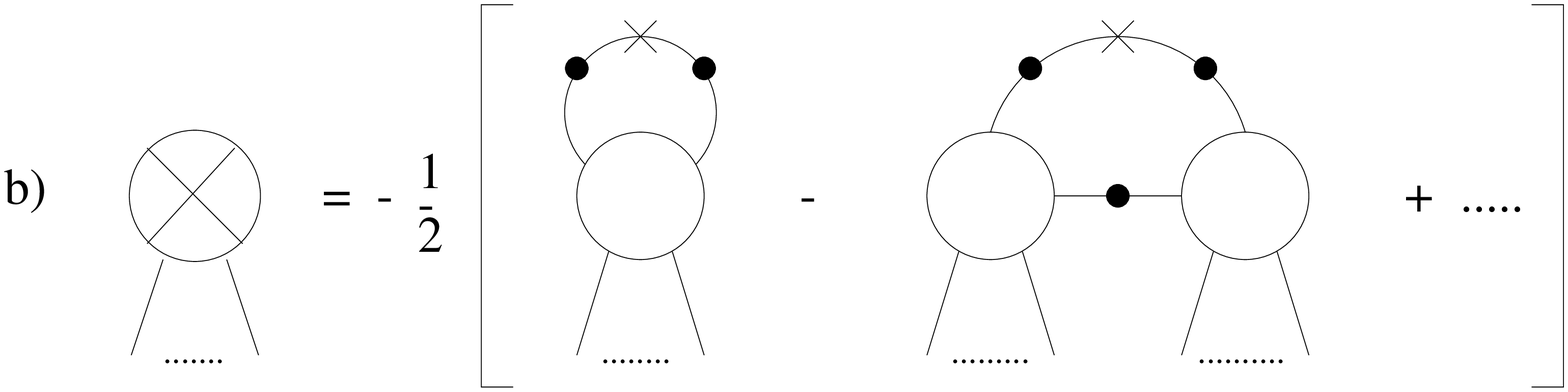,height=12 ex}
    \end{tabular}
  \end{center}
  \label{1}
  \caption{{\small
Graphical representation of the \lRG equations for the two point $(a)$ 
and $2n$-point $(b)$ vertex functions.
The black dot represents the full propagator; open circles represent 
vertex functions with the IR cutoff $\L$; crosses represent 
the derivative with respect to $\L$ of the free 
cutoff propagator and vertices. 
Integration over $q$ in the loop is understood.}}
\end{figure}
The RG equations \re{EvEq} have to be supplemented by boundary 
conditions at some values of $\L$. To discuss them one observes that
at the UV scale $\L=\L_0$ the Wilsonian or cutoff effective action
$\G[\phi;\L,\L_0]$ reduces to the UV action \re{S0}
\beq\label{bc1}
\G[\phi;\L_0,\L_0]=S_{\L_0}[\phi]
\,.
\eeq
Moreover, when the IR and UV cutoff are removed 
($\L=0$ and $\L_0\to\infty$)
and the UV parameters $\rUV_i$ have the proper dependence on $\L_0$,
the cutoff vertices $\G_{2n}(\{p_i\};\L,\L_0)$ become
the physical vertices $\G_{2n}(\{p_i\})$
\beq\label{bc2}
\G_{2n}(\{p_i\})= \G_{2n}(\{p_i\};\L=0,\infty)
\,.
\eeq
The boundary conditions for the RG evolution equations are therefore
given by the requirement of locality \re{bc1}, \ie by giving the values of the
``relevant'' UV parameters $\rUV_i$ in \re{S0} and by requiring that all
parts of vertices with negative dimension vanish.
Without knowing the elementary theory at distances shorter than
$1/\L_0$, the UV parameters $\rUV_i$ must be obtained from
measurement at low scale. 
To define this procedure one has to introduce in general, for any
$\L$, the relevant parameters together with their irrelevant parts. 

\noindent
{\it Relevant parameters, subtraction point and physical couplings}.
In the four dimensional massless scalar $\phi^4$ theory the two and four-point
vertices $\G_2$ and $\G_4$ have non-negative mass dimensions and one
has three relevant parameters defined, for instance, as follows
\beq\label{relev}
\eqalign{
&
\G_2(p;\L,\L_0)=D^{-1}_{\L\L_0}(p)+p^2\;(Z(\L)-1)\;+\;\L^2\r_1
(\L)\;+\;\Delta_2(p;\L,\L_0) \,,
\cr&
\G_4(p_1,p_2,p_3,p_4;\L,\L_0)
=\r_2(\L)+\Delta_4(p_1,p_2,p_3,p_4;\L,\L_0)
\,,
}
\eeq
with the conditions
\beq\label{delta}
\Delta_2(0;\L,\L_0)
=\frac {d\Delta_2(p;\L,\L_0)}{dp^2}\vert_{p^2=\mu^2}=0
\,,\;\;\;\;\;\;\;\;
\Delta_4(\bp_1,\bp_2,\bp_3,\bp_4;\L,\L_0)=0
\,,
\eeq
where $\mu$ is some subtraction scale and $\bp_i$ is the
symmetric point $\bp_i\bp_j=\mu^2(\delta_{ij}-\frac14)$.
At the physical point, from \re{bc2} one has
\beq\label{bc3}
Z(0)=1\,,\;\;\;
\L^2\r_1(\L)|_{\L=0} =0\,,\;\;\;\;
\r_2(0) =g\,,
\eeq
for large $\L_0$.
These conditions set the field normalization, the mass 
equal to zero and the coupling at $\mu$.

The relevant parameters at the physical point $\L=0$ 
are related by the \lRG equations to the UV parameters $\rUV_i$.
Therefore as boundary conditions  one can take for the relevant 
parameters the physical values \re{bc3} and the vanishing of the 
irrelevant vertices at the UV point (locality condition \re{bc1}).
In Appendix we recall how the loop expansion can be generated from the
RG equations for the cutoff vertices $\G_{2n}(\{p_i\};\L,\L_0)$
with these boundary conditions.

For the relevant parameters $\r_i(\L)$ and $Z(\L)$ we have explicitly 
written only the $\L$ dependence. However, from their definition in
\re{relev} they are dimensionless functions of $\L$, $\L_0$ and $\mu$. 
One should write $ \r_i(\L) \equiv \r_i(g,\mu/\L,\mu/\L_0)$ and 
$Z(\L) \equiv Z(g,\mu/\L,\mu/\L_0)$.
In perturbation theory one can take the limit $\L_0\to\infty$ and
all physical vertices become $\L_0$ independent.
This implies that one has for large $\L_0$
\beq\label{ri'}
\r_i(\L) \;\equiv\; \r_i(g,{\mu}/{\L}, {\mu}/{\L_0})
\;=\;\r_i(g,{\mu}/{\L},0) \;+\;{\cal O} ({\mu}/{\L_0})
\,,
\eeq
and similarly for $Z(\L)$.
This allows one to simplify the perturbative analysis
by putting $\L_0\to \infty$ from the beginning (see ref.~\cite{BDM}).
However, in the scalar $\phi^4$ theory, non-perturbative triviality
allows one to take this limit only for vanishing $g$.

\subsection{Relation between \lRG and \mRG}
As well known renormalization is related to the UV property of the
theory. One should then expect that \mRG, 
which corresponds to the requirement that the physical
observables do not depend on the specific value of $\mu$,
is related to \lRG for $\L$ in the UV region.
Actually, as we will show, \mRG corresponds to renormalizability,
\ie the independence of physical vertices on the UV cutoff $\L_0$
for large $\L_0$. The discussion on this point will be then in the
framework of perturbation theory.

From \re{bc1} for $\L$ approaching the UV point $\L_0$
the cutoff effective action $\G[\phi;\L,\L_0]$ becomes the local 
UV action \re{S0} and one can write
\beq\label{UVlim}
\eqalign{
&
\G[\phi;\L_0,\L_0]= S_{\L_0}[\phi]
=\int d^4x
\left\{
\half
\phi^\U(x) \;[-\partial^2+\L_0^2\;\brUV_1]\;\phi^\U(x)
\;+\;\frac{\brUV_2}{4!} (\phi^\U(x))^4
\right\}
\,,
\cr&
\phi^\U=\sqrt{Z^\U}\phi
\,,
\;\;\;\;\;\;
\brUV_1=\frac{\rUV_1}{Z^\U}\,,
\;\;\;\;\;\;
\brUV_2=\frac{\rUV_2}{(Z^\U)^2}
\,.
}
\eeq
The UV parameters are given by 
\beq\label{rUV}
\rUV_i \equiv
\r_i(g,{\mu}/{\L},{\mu}/{\L_0})|_{\L=\L_0}
\,,\;\;\;\;
Z^\U=Z(g,\mu/\L,\mu/\L_0)|_{\L=\L_0}
\,.
\eeq
Consider now the physical vertices $\G_{2n}(\{p_i\},g,\mu)$
obtained from this UV action.
The rescaling of the fields in \re{UVlim} allows one to explicitly
factorize the dependence on $Z^\U$
\beq\label{GUV}
\G_{2n}(\{p_i\},g,\mu)=
\left(Z^\U\right)^n\;
\G_{2n}^\U(\{p_i\}, \brUV_1,\brUV_2;\L_0)
\,.
\eeq
These UV vertices $\G^\U_{2n}$ are obtained from \re{UVlim} by Feynman
rules with internal propagators of momentum $q$ with the UV cutoff
$q^2<\L_0^2$.
The various Feynman diagram contributions to
$\G^{\mbox{\scriptsize UV}}_{2n}$
contain logarithmic divergent powers of
\beq\label{logsplit}
\ln\left(\frac{P^2}{\L^2_0}\right)
=\ln\left(\frac{P^2}{\mu^2}\right)
+\ln\left(\frac{\mu^2}{\L_0^2}\right)
\,,
\eeq
with $P$ a combination of external momenta.
The two point vertex contains also quadratically divergent mass terms
proportional to $\L_0^2$.
Renormalizability ensures that (perturbatively) all these $\L_0^2$
divergences are cancelled by the divergent contributions in the UV
parameters $\rUV_i$ and $Z^\U$ so that for large $\L_0$ the vertex functions
are independent of $\L_0$. This fact is expressed by the equation
\beq\label{lRG}
\left[
\L_0\partial_{\L_0}
+(\L_0\partial_{\L_0}\brUV_1)\frac{\partial}{\partial \brUV_1}
+(\L_0\partial_{\L_0}\brUV_2)\frac{\partial}{\partial \brUV_3}
+n\L_0\partial_{\L_0}\ln Z^\U
\right]\G_{2n}(\{p_i\},g,\mu)=0
\,,
\eeq
valid for large $\L_0\to \infty$, \ie the corrections are of order
$\mu/\L_0$. Here $\partial_{\L_0}$ is the explicit derivative with
respect to $\L_0$ in \re{GUV}.

To relate this equation to the \mRG equation one needs to analyse the
$\mu$ dependence of the vertices and then of the UV parameters. 
The key observation is that, for $\L_0\to\infty$, the complete $\mu$
dependence in \re{GUV} is contained only in the factor $(Z^\U)^n$.
This is due to the fact that the physical vertices
$\G_{2n}(\{p_i\},g',\mu')$ with coupling $g'$ at a scale $\mu'$
are related to the ones with coupling $g$ at $\mu$ by the relation
\beq\label{Z2n}
\G_{2n}(\{p_i\},g',\mu')=z^{-n}\; \G_{2n}(\{p_i\},g,\mu)
\,,
\;\;\;\;\;\;
z=z(g,{\mu'}/{\mu})= Z^\U/Z'^{\U}
\,,
\eeq
with $Z'^\U$ given in \re{rUV} as a function of $g'$ and $\mu'$. 
Therefore the UV parameters $\brUV_1$ and $\brUV_2$ become independent 
of $\mu$ for $\L_0 \to\infty$. In this limit one obtains the following 
relation between beta function and UV parameters 
\beq\label{betal}
\beta(g) \equiv
\mu\partial_{\mu}g
\;=\;\frac {\L_0\partial_{\L_0} \, \brUV_2}{\partial_g \, \brUV_2}
\;=\;\frac {\L_0\partial_{\L_0} \, \brUV_1}{\partial_g \, \brUV_1}
\,.
\eeq
For the field anomalous dimension one has 
\beq\label{gammal}
\g_\phi(g)
= -\half \mu'\partial_{\mu'}\;\ln z(g,\mu'/\mu)|_{\mu'=\mu}
=
- \half [\L_0\partial_{\L_0}-\beta(g)\partial_g] \; \ln Z^\U
\,.
\eeq
Notice that the UV parameters are logarithmically divergent
for $\L_0\to\infty$,
but (perturbative) renormalizability ensures that, in this limit,
$\beta(g)$ and $\g_\phi(g)$ are finite and dependent only on $g$.

It is easy to show that the renormalizability condition \re{lRG} is 
equivalent to the \mRG equation. From \re{betal} and \re{gammal} 
we can write \re{lRG} in the form ($\L_0\to\infty$)
\beq\label{lRG1}
\eqalign{
&
\left[
\L_0\partial_{\L_0}
+\beta(g)
\left(
 \partial_g\brUV_1\frac{\partial}{\partial \brUV_1}
+\partial_g\brUV_2\frac{\partial}{\partial \brUV_2}
+\partial_g Z^\U\frac{\partial}{\partial Z^\U}
\right)
-2n\g_\phi(g)
\right]
\G_{2n}(\{p_i\},g,\mu)
\cr&
=\left[ \L_0\partial_{\L_0} +\beta(g) \partial_g -2n\g_\phi(g)
\right]\G_{2n}(\{p_i\},g,\mu)=0
\,.
}
\eeq
From \re{logsplit} we see the explicit logarithmic $\L_0$
dependence is the same as the explicit $\mu$ dependence, thus
$\L_0\partial_{\L_0}$ is equivalent to $\mu\partial_{\mu}$ and one 
recovers the usual \mRG equation.

The fact that the theory is (perturbatively) renormalizable ensures
that the $\L_0$ dependence of the relevant parameters vanishes for
$\L_0\to \infty$ (see \re{ri'}).
This implies that we can introduce the relevant parameters
$\r_i(g,\mu/\L,\mu/\L_0)$ directly for $\L_0\to\infty$. 
From now on we shall use
$$
\r_i(\L)=\r_i(g,{\mu}/{\L},0)
\,,\;\;\;\;
Z(\L)=Z(g,{\mu}/{\L},0)
\,,
$$
so that the beta function in eq.~\re{betal} can be expressed in
terms of these functions for large $\L$, \ie $\mu/\L\to 0$.
For the beta function and the anomalous dimension one has
\beq\label{betagamma}
\eqalign{
&
\beta(g)
=\left[\frac{\L\partial_{\L}\,\br_1}{\partial_g\,\br_1}\right]_{\mu/\L=0}
=\left[\frac{\L\partial_{\L}\,\br_2}{\partial_g\,\br_2}\right]_{\mu/\L=0}
\,,
\cr&
\g_\phi(g)
=-\half [\L\partial_{\L} -\beta(g)\partial_g] \; \ln Z|_{\mu/\L=0} \,,
}
\eeq
with $\br_2=\r_2/Z^2$ and $\br_1 = \r_1/Z$.
The parameters $\br_i$ have logarithmic divergent contributions for
$\mu/\L \to 0$ and only the ratios in \re{betagamma} are finite.
The relation \re{betagamma} between the beta function and the
parameters $\br_2$ was obtained in Ref.~\cite{H} to one-loop order
where the denominator $\partial_g \br_2$ can be ignored.
As we shall see at two loops the denominator is essential to obtain a
finite result for $\mu/\L \to 0$.

In the following we shall focus our attention on $\br_2$ which will be
called the ``Wilsonian flowing coupling'', $\tg(\L)\equiv \br_2(\L)$.
We show now that for large $\L$ this coupling becomes equivalent to a
running coupling at the momentum scale $k^2=\L^2\gg \mu^2$.
This is a consequence of the fact that for large $\L$ this coupling
becomes independent of $\mu$ so that one can write
$$
\tg(\L)  \equiv \br_2(g,\mu/\L)
=G(g(\mu),\mu/\L)\;(1\;+\; {\cal O}(\mu/\L))
\,,
$$
where the function $G$ does not depend on $\mu$.
This means that, apart for power corrections, the $\mu$ dependence in 
$g(\mu)$ is completely cancelled by the explicit $\mu/\L$ dependence. 
We then can write
\beq
\tg(\L)=G(g(\L),1)\;(1\;+\; {\cal O}(\mu/\L))
\,,\;\;\;\;\;
G(g(\L),1)=g(\L) + {\cal O}(g^2(\L))
\,,
\eeq
so that $G(g(\L),1)$ itself is a possible running coupling at the scale $\L$.

\subsection{Beta function at two loops by \lRG}
To illustrate in some details the \lRG formulation we compute the two 
loop beta function and anomalous dimension using \re{betagamma}
with $\br_2$. We have to compute the behaviour for $\mu/\L \to 0$ 
of the perturbative coefficients of $\r_2$ and $Z$ 
\beq
\r_2(g,\mu/\L)= g+ \sum_{\ell=1}^\infty
\;g^{\ell+1}\;\r^{(\ell)}_2(\mu/\L) \,,
\;\;\;\;\;
Z(g,\mu/\L)=1+\sum_{\ell=2}^\infty \;g^\ell\; Z^{(\ell)}(\mu/\L)
\,.
\eeq
The corrections to $Z$ start from two loops since the one-loop self 
energy does not depend on the momentum.
The first two coefficients of the beta function
\beq
\beta(g)=b_0 g^2 +b_1 g^3+ \cdots
\eeq
are given by (see \re{betagamma}) 
\beq\label{beta2}
b_0=\dr_2^{(1)}|_{\mu=0}
\,,
\;\;\;\;
b_1=\;(\,\dr_2^{(2)} -2\dr_2^{(1)}\r_2^{(1)} -2\dZ^{(2)}\;)|_{\mu=0}
\,,
\;\;\;\;\;
\dr_i(\L) \equiv \L\partial_{\L}\r_i(\L)
\,.
\eeq
As we shall see, in the expression for $b_1$ the divergence contribution 
in $\dr_2^{(2)}$ is cancelled by the divergent contribution 
$-2\dr_2^{(1)}\r_2^{(1)}$, which is coming from $\partial_g\br_2$ in 
the denominator of \re{betagamma}.

At one loop the relevant parameters are obtained from fig.~1
with the vertices at tree level. 
The momentum dependence of the self energy vanishes ($Z^{(1)}=0$).
Neglecting the subtraction point one finds
\beq
b_0=\dr_2^{(1)}|_{\mu=0}
=-3 \int_{q}\frac{\dot K_\L(q)}{q^2}\frac{K_\L(q)}{q^2}
=-\frac 3 2  \int_{q} \frac {1}{q^4} \L \partial_\L K^2_\L(q)
= 3r\,, \;\;\;\;\;\; r=\frac1{16\pi^2}
\,.
\eeq
Notice that since $\dr_2^{(1)}(\mu/\L)$ is a constant for $\mu/\L=0$,
$\r_2^{(1)}(\mu/\L)$ diverges logarithmically as $\ln \mu/\L$.

The two-loop coupling $Z^{(2)}(\mu/\L)$ is obtained from fig.~1a by 
taking the four point function at one-loop (to this order self 
energy corrections
in internal propagators do not contribute).
\begin{figure}
  \begin{center}
    \mbox{\epsfig{file=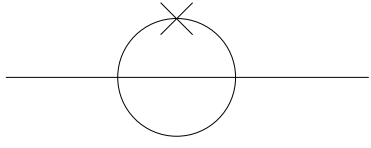,height=10 ex}}
  \end{center}
  \label{2}
  \caption{{\small Feynman diagram contributing to $\dZ^{(2)}(\L)$.}}
\end{figure}
Neglecting the subtraction point $\mu$ one has (see fig.~2)
\beq\label{Z2}
\dZ^{(2)}|_{\mu=0}
=
-\frac 16 \int_{q}\int_{q'} \L\partial_\L \;\partial_{p^2}
\left\{
\frac{K_\L(q) K_\L(q') K_\L(q+q'+p) }{q^2q'^2(q+q'+p)^2}
\right\}_{p=0}
=- \frac{r^2}{6}
\,.
\eeq
These integrals are easily done by neglecting one of the cutoff function.
The remaining contribution gives a constant which vanishes when the
$\L$-derivative is taken.

The two-loop coupling $\dr_2^{(2)}(\mu/\L)$ is obtained from the
diagrams of fig.~3a, in which the internal vertices are at one loop, and
of fig.~3b. 
\begin{figure}
  \begin{center}
    \begin{tabular}{c}
    \epsfig{file=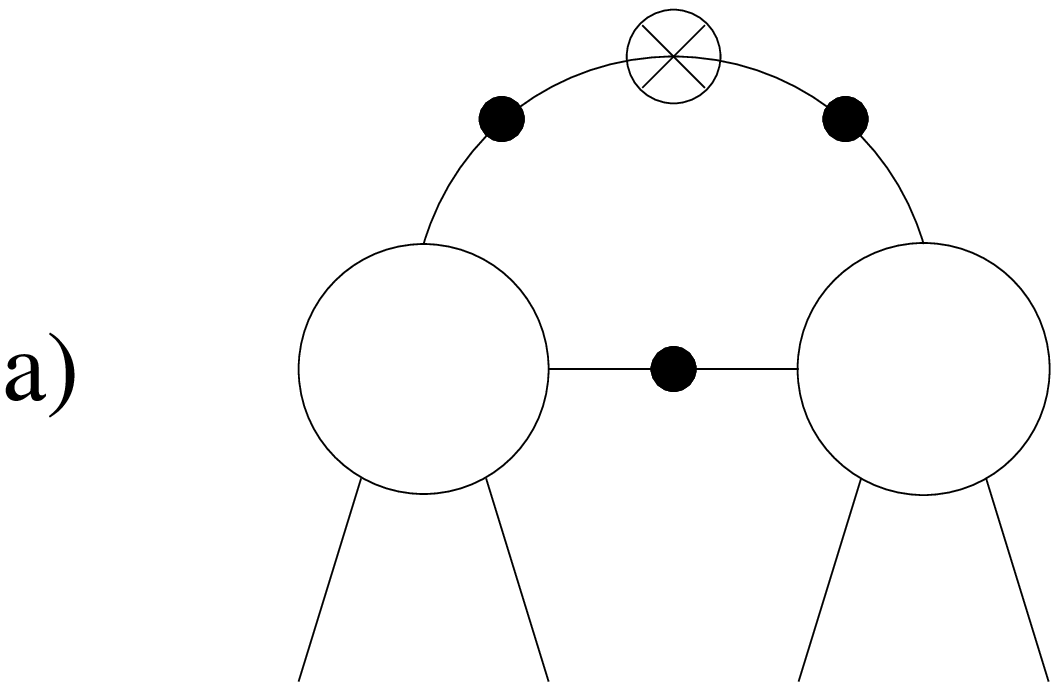,height=12 ex}\\
     {}\\
    \epsfig{file=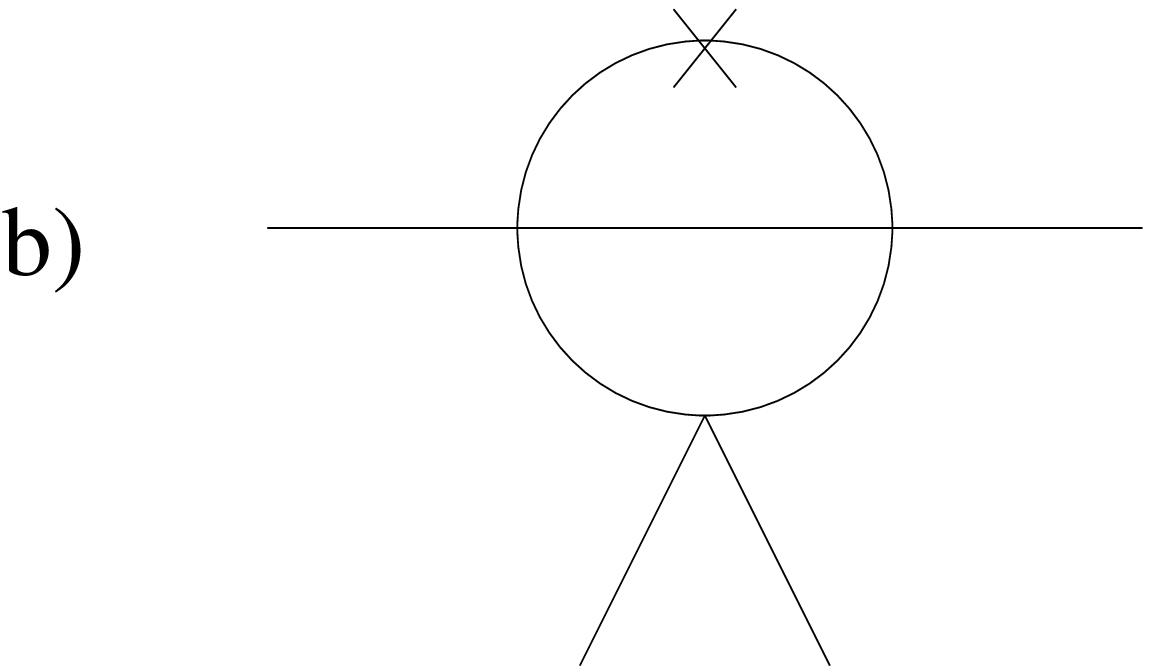,height=12 ex}
    \end{tabular}
  \end{center}
  \label{3}
 \caption{
{\small $(a)$ Graphical representation of the contribution to $\dr_2$ 
(see the second term of \re{I6}). 
$(b)$ Feynman diagram contributing to $\dr_2^{(2)}$ (see
the first term of \re{I6}). 
}}
\end{figure}
The first contribution is
\beq\label{rA}
g^3\dr_{2A}^{(2)}|_{\mu=0} =
-\frac 32 \; \int_q
\left[
\G^2_4(q,0,0,-q;\L)
\; \L\partial_\L
\;\frac{1}{\G^2_2(q;\L)}
\right]_{\mbox{one loop}}
\,.
\eeq
We have to consider the three contributions
$\r_1^{(1)}(\L)$,
$\r_2^{(1)}(\L)$ and
$\Delta_4^{(1)}(q,0,0,-q;\L)$
coming from the one-loop corrections to the two- and four-point functions.

The contribution from  $\r_1^{(1)}(\L)$ is zero. This is due
to the fact that, before taking the $\L$ derivative, the integral is
convergent, dimensionless and then its result is a constant in $\L$.

The contribution from  $\r_2^{(1)}(\L)$ is
\beq\label{rA'}
\dr_{2A'}^{(2)}|_{\mu=0} =
-3 \;\r_2^{(1)}
\;
\int_q
\; \L\partial_\L
\;\left(\frac{K_{\L}(q)}{q^2}\right)^2
= 2\;\r_2^{(1)}\;\dr_2^{(1)}
\,.
\eeq
As anticipated, this contribution is cancelled by the term 
$-2\;\r_2^{(1)}\;\dr_2^{(1)}$ in the combination \re{beta2} which 
is coming from the denominator $\partial_g \br_2$.

The contribution from  $\Delta_4^{(1)}(q,0,0,-q;\L)$ is
\beq\label{rA''}
\eqalign{
\dr_{2A''}^{(2)}|_{\mu=0}
&
=3 \;\int_q \;\L\partial_\L\; \left(\frac{K_\L(q)}{q^2}\right)^2
\;\int_{q'}\frac{K_\L(q')}{q'^2}\;
\left(\frac{K_\L(q+q')}{(q+q')^2}-\frac{K_\L(q')}{q'^2}\right)
\cr&
=3 \;\int_q \;\L\partial_\L\; \left(\frac{K_\L(q)}{q^2}\right)^2
\;\int_{q'}\frac{K_\L(q')\;(K_\L(q+q')-1)}{q'^2(q+q')^2}
\,,
}
\eeq
where the equality of the two integrals can be shown by performing the
angular integration.

Finally the contribution from the diagram of fig.~3b is
$$
\dr_{2B}^{(2)}|_{\mu=0}
=3 \int_{q'}\left(\frac{K_\L(q')}{q'^2}\right)^2
\;\int_q \;\L\partial_\L\; \frac{K_\L(q)K_\L(q+q')}{q^2(q+q')^2}
\,.
$$
It is easy to compute this integral if one of the cutoff functions inside 
the second integral is neglected 
\beq\label{rB'}
\dr_{2B'}^{(2)}|_{\mu=0}
=3 \int_{q'}\left(\frac{K_\L(q')}{q'^2}\right)^2
\;\int_q \;\L\partial_\L\; \frac{K_\L(q)}{q^2(q+q')^2}
= -6 r^2 
\,.
\eeq
The remaining term cancels exactly the contribution \re{rA''}.

In conclusion the only non-cancelling contributions to
the two loops beta function and anomalous dimension are given by
\re{Z2} and \re{rB'} and one finds
\footnote{
In Ref.~\cite{Mo} an approximate perturbative calculation of the 
beta function has been obtained using a momentum expansion and a 
truncation of the RG equations which omits 
the contribution of the six point vertex.}
\beq\label{2loops}
b_1 =-\frac{17}{3} r^2 
\,,
\;\;\;\;\;
\g^{(1)}_\phi(g)=\frac{1}{12} g^2 r^2\,.
\eeq
The evaluations given above are obtained essentially by using a theta
function for the cutoff function, but the results do not depend on
this specific assumption and depend essentially on dimensional
counting of convergent integrals. This is due to the fact that one 
takes the limit $\L\to\infty$ (a smooth representation of the theta 
function gives corrections vanishing as inverse powers of $\L$). 

\subsection{ Flowing coupling and resummation}
As recalled in the Introduction, it is usually assumed that in a
Feynman diagram with a virtual loop of momentum $k$ higher order 
corrections reconstruct the running coupling $g(k)$ at that 
momentum scale. 
We want here to show that the \lRG formulation may provide a simple
framework in which one can study this resummation.

As we have seen in the previous subsections (see also the Appendix) 
the usual perturbative expansion is obtained by solving iteratively 
the \lRG evolution equations with the initial condition that at zero 
loop the only non vanishing vertices are
$
\G^{(0)}_4(p_1 \cdots p_4;\L)=g
$
and $\G^{(0)}_2(p;\L)=D^{-1}_{\L\L_0}(p)$.
By iterating the \lRG equations $\ell$ times one generates 
contributions of Feynman diagrams at $\ell$ loops in which 
the loop momenta $k_i$ are ordered in virtuality 
$\L^2<k^2_1<\cdots <k^2_\ell$. The complete
Feynman diagrams at that loop are obtained by summing all orderings.
This ordering is due to the fact that $\L$ is an IR cutoff $\L^2<k_1^2$ 
and then in the next iteration $k_1^2$ plays the r\^ole of an IR cutoff.

It is possible to formulate an improved iterative procedure in which 
one introduces the Wilsonian flowing couplings $\tg(k_i)$ 
at the scale $k_i$ of the successive momenta generated by the iteration.
This improved perturbative expansion consists in solving the \lRG 
equations as for the usual perturbative expansion with the only 
difference that at zero loop the four point function is given by 
the flowing coupling 
\beq\label{zero}
\G^{(0)}_4(p_1 \cdots p_4;\L)
=\tg(\L)
\,.
\eeq
The flowing coupling will be obtained later by a generalized beta 
function computed as an expansion in the Wilsonian coupling itself.

From \re{zero} one starts the iteration by computing to the next loop 
the irrelevant part of the vertices and the two relevant parameters 
$Z(\L)$ and $\r_1(\L)$ as function of the flowing coupling. 
After iterating this procedure $\ell$ times, the vertices are given in 
terms of flowing couplings $\tg(k_i)$ at the scale $k_i$ of the 
successive momenta generated by the iteration.

From the vertices obtained after $\ell$ iterations one computes 
the generalized beta function 
\beq\label{Gbeta}
B[\tg;\L] \equiv \L\partial_\L \tg(\L)
=-2 \frac{\dot Z}{Z} \;\tg(\L)
+ \half \;\int_q\;Z^{-2}(\L)\;\bar I_6(q,\bp_1\cdots \bp_4,-q;\L)
\;\L\partial_\L \frac{1}{\G_2(q;\L)}
\,.
\eeq
The vertex $\bar I_6$ is defined in the Appendix in terms
of the two- four- and six-point cutoff vertices.
Solving \re{Gbeta} with the boundary condition $\tg(0)=g$, one 
computes $\tg(\L)$ as a partial resummation of perturbative expansion. 
Notice that $B[\tg;\L]$ is a functional of the flowing coupling 
$\tg(k_i)$ and the scale $\L$. 
This has to be contrasted with $\beta(g)$ which is a function of $g$.

To illustrate the procedure we compute the flowing coupling to one-loop
order (similar approximation has been considered for large $\L$ in 
Ref.~\cite{LP}). This is obtained by taking the vertices in the integrand 
of \re{Gbeta} to zero-loop
\beq
\bar I^{(0)}_6(q,p_1\cdots  p_4,-q;\L)=
-2\tg^2(\L)\;\left[
D_{\L\L_0}(Q_{12})+ D_{\L\L_0}(Q_{13})+D_{\L\L_0}(Q_{14})
\right]
\,,
\eeq
where $Q_{ij}=q+p_i+p_j$. The equation for $Z(\L)$ gives to this order 
$\dot Z^{(1)}(\L)=0$ and one obtains
$$
B^{(1)}[\tg;\L]=b_0\;\tg^2(\L)\;f(\mu/\L)
\,,
$$
where $b_0$ is the one-loop coefficient of the beta function and
\beq\label{f2}
f(\mu/\L) = -8\pi^2 \int_q\; \L\partial_\L
\;\frac{K_{\L}(q)\,K_{\L}(q+\bp)}{q^2(q+\bp)^2}
\;=\;1-\frac{2\mu}{3\pi\L}+{\cal O}(\mu^2/\L^2)
\,,\;\;\;\;\;\L\to\infty
\,.
\eeq
For $\L\to 0 $ this function vanishes and one finds 
$f(\mu/\L)=2 (\L/\mu)^2 +{\cal O}(\L/\mu)^3$.

To this order $\tg(\mu/\L)$ is obtained by solving
\beq
\L\partial_\L \tg(\L) = b_0\;\tg^2(\L)\;f(\mu/\L)
\,,
\;\;\;\;\; \tg(0)=g\,.
\eeq
As expected, since $f(\mu/\L)\to 1$ for large $\L$, the flowing and
running coupling become the same for large $\L$
\beq
\tg(\L) = \frac{g}{1-b_0g \ln(\L/\mu_0)}
\left(1 +{\cal O}(\frac{\mu}{\L})\right)
\,,
\eeq
where $\mu_0\simeq\mu$.
For smaller $\L$ the two couplings
differ: $\tg(0)=g$ while 
$g(\mu)=g$ and $g(k) \to 0$ for $k \to 0$.

The physical vertices are obtained by integrating the virtualities
between $\L=0$ and $\L_0$. The presence of the Landau pole at
$k_L \simeq \mu \,e^{1/b_0g}$ implies that the UV cutoff cannot be
removed in this case, although the theory is perturbatively
renormalizable.
This corresponds to the property of triviality of the scalar theory,
entailing that the limit $\L_0 \to \infty$ is possible
only if the coupling $g$ vanishes.

\section{Yang-Mills case}

We show that the previous connection between \mRG and \lRG
formulations can be established also for Yang-Mills theory and we
shall compute as an example the one-loop beta function.
We recall that the \lRG for this theory is consistent with
the BRS symmetry \cite{BRS} provided the boundary conditions at the
physical point $\L=0$ and $\L\to\infty$ are properly chosen
\cite{B,BDMYM} (see also \cite{others}).
We then consider the iterative solution in terms of the Wilsonian
couplings.

\subsection{Summary of \lRG formulation}
Consider as an example the $SU(2)$ Yang-Mills theory
in which $\phi=(A_\mu^a,c^a,\bc^a)$ are the vector and the ghost fields
and $\g=(u_\mu^a,v^a)$ the sources for composite operators
associated to the BRS variations of $A_\mu^a$, and $c^a$ respectively.
The BRS classical action, in the Feynman gauge, is
\beq\label{BRS}
S_{\mbox{\footnotesize{BRS}}}
=\int d^4x \biggl\{-\frac 1 4 F_{\mu\nu}^2
-\frac 1 2 (\partial_\mu A_\mu)^2
+ W_\mu \cdot D_\mu c - \frac 1 2 \, v\cdot c \wedge c
\biggr\}\,,
\eeq
with $F_{\mu\nu}=\partial_\mu A_\nu- \partial_\nu A_\mu +
g A_\mu \wedge A_\nu$, $D_\mu c =\partial_\mu c +gA_\mu \wedge c$,
$W_\mu=\partial_\mu \bc+u_\mu/g$ and we have introduced
the usual scalar and external $SU(2)$ products.

In the \lRG formulation one introduces the cutoff effective action
$\G[\phi,\g;\L,\L_0]$ by using propagators with IR and UV
cutoff at $\L$ and $\L_0$ respectively and one deduces the RG 
equation similar to \re{EvEq}. Then one has to fix the boundary conditions.
First of all one assumes that at the UV point $\L=\L_0$ the cutoff
effective action becomes local.
Then one has to fix the relevant parameters of $\G[\phi,\g;\L,\L_0]$,
\ie the coefficients of all independent fields monomials which have
dimension not larger than four and are Lorentz and $SU(2)$ scalars.
There are nine relevant parameters.
Two are the wave function constants $Z_A(\L)$ and $Z_c(\L)$ for
the vector and ghost fields respectively which are the
coefficients of the monomials
$$
\half A_\mu (g_{\mu\nu}\partial^2-\partial_\mu\partial_\nu)   
\cdot A_\nu\,,
\;\;\;\;\;
W_\mu\cdot \partial_\mu \;c
\;.
$$
The remaining parameters $\r_i(\L)$ with $i=1\cdots7$ can be taken as the
coefficients of the following seven monomials
$$
\half \L^2 \, A_\mu \cdot A_\mu
\,,\;\;\;\;
-\half (\partial_\mu A_\mu)^2
\,,\;\;\;\;
(\partial_\nu A_\mu) \cdot A_\mu \wedge A_\nu
\,,
$$
$$
W_\mu \cdot A_\mu \wedge c
\,,
\;\;\;\;
\frac 14 (A_\mu\wedge A_\nu)\cdot(A_\nu\wedge A_\mu)
\,,\;\;\;\;
\frac 14 (A_\mu\cdot A_\mu)^2
\,,\;\;\;\;
-\half v\cdot c \wedge c
\,,
$$
respectively.
Notice that some of these monomials are not present in the classical
action \re{BRS}.
As in the scalar case (see \re{relev},\re{delta}) these nine parameters 
are given in terms of vertex functions evaluated at the
subtraction points at the scale $\mu$.

At the physical point $\L=0$ and $\L_0\to\infty$ the effective action 
$\G[\phi,\g;0,\infty]$ satisfies the BRS symmetry. 
As shown in Refs.~\cite{B,BDMYM} this symmetry is
ensured if one properly fixes the relevant parameters defined at a
given subtraction point $\mu$ as boundary conditions at $\L=0$. 
Taking the relevant part of the effective action, \ie the part 
which contains only the relevant parameters, one has
\beq\label{bcBRS}
\Gr[\phi,\g;0,\infty]
=S_{\mbox{\footnotesize{BRS}}}
+\hbar
\int d^4x \biggl\{
 \frac{x_5}{4}(A_\mu\wedge A_\nu)\cdot(A_\nu\wedge A_\mu)
+\frac{x_6}{4} (A_\mu\cdot A_\mu)^2
-\half{x_7} \, v\cdot c \wedge c
\biggr\}
\,,
\eeq
where the three parameters $x_i$ are given in terms of 
irrelevant vertices evaluated at the subtraction points (see
\cite{BDMYM}).
Therefore as boundary conditions at the physical 
point $\L=0$ and $\L_0\to\infty$ one assumes
$$
Z_A(0)=1
\,,\;\;\;\;
Z_c(0)=1
\,,\;\;\;\;
\r_1(0)=0
\,,\;\;\;\;
\r_2(0)=1
\,,\;\;\;\;
\r_3(0)=g
\,,\;\;\;\;
\r_4(0)=g
\,,
$$
$$
\r_5(0)=g^2+\hbar x_5
\,,\;\;\;\;
\r_6(0)=\hbar x_6
\,,\;\;\;\;
\r_7(0)= 1 + \hbar x_7
\,.
$$
With these conditions the \lRG equation can be solved by expanding
in $\hbar$ and one obtains the usual perturbative expansion
for the vertex functions which satisfy the Slavnov-Taylor identities.

\subsection{Beta function in the \lRG formulation}

As in the scalar case we consider the cutoff effective action
at the UV point.
For $\L=\L_0$ the functional $\G[\phi,\g;\L,\L_0]$ becomes local
and equal to the UV action $S_{\L_0}[\phi,\g]$.
As in \re{rUV} we introduce the UV wave function constants and
UV couplings
$$
Z_i^\U=Z_i(g,\mu/\L,\mu/\L_0)|_{\L=\L_0}
\,,\;\;\;\;
\r_i^\U=\r_i(g,\mu/\L,\mu/\L_0)|_{\L=\L_0}
\,.
$$
Rescaling the fields, the UV action is given by
\beq\label{UVYM}
\eqalign{
&
\G[\phi,\g;\L_0,\L_0]
=S_{\L_0}[\phi,\g]
\cr&
=\int d^4x \biggl\{
\half A^\U_\mu (g_{\mu\nu}\partial^2-\partial_\mu\partial_\nu)
\cdot A^\U_\nu
\;+\; W^\U_\mu\cdot\partial_\mu c^\U
\;-\; \half v^\U\cdot c^\U\wedge c^\U
\cr&
\;+\;\half\L_0^2 \,\br^\U_1 A^\U_\mu \cdot A^\U_\mu
\;-\;\half \br^\U_2 (\partial_\mu A^\U_\mu)^2
\;+\;\br^\U_3 (\partial_\nu A^\U_\mu) \cdot A^\U_\mu \wedge A^\U_\nu
\cr&
+\br^\U_4 W^\U_\mu \cdot A^\U_\mu \wedge c^\U
\;+\;\frac{\br^\U_5}{4} (A^\U_\mu\wedge A^\U_\nu)\cdot(A^\U_\nu\wedge
A^\U_\mu) +\frac{\br^\U_6}{4} (A^\U_\mu\cdot A^\U_\mu)^2
\biggr\}
\,,
}
\eeq
where
\beq
A_\mu^\U=\sqrt{Z_A^\U}\;A_\mu
\,,\quad\quad
W^\U_\mu\, c^\U=Z_c^\U\;W_\mu \,c
\,.
\eeq
The UV parameters $\br_i^\U$ are obtained as in \re{UVlim}
by the usual rescaling. For instance one has
$$
\br_3^\U =\r_3^\U \;(Z_A^\U)^{-3/2}  \,,\;\;\;\;\;
\br_4^\U =\r_4^\U \;(Z_c^\U)^{-1} \;(Z_A^\U)^{-1/2} \,,\;\;\;\;\;
\br_5^\U =\r_5^\U \;(Z_A^\U)^{-2} 
\,.
$$
The BRS source $v^\U$ is rescaled with no other effect.
In the following we consider only vertex functions involving 
the vector and ghost fields and neglect the $v^\U$ source.

The beta function and the field anomalous dimensions are obtained as
in the scalar case.
After factorizing the UV wave functions (see \re{GUV}) the
vertices depend only on the six UV parameters $\br_i^\U$ with
$i=1,\cdots 6$.
Since the $\mu$ dependence is in the factorized wave function
$Z^\U_i$ one deduces that the six UV rescaled parameters $\br_i^\U$
do not depend on the subtraction point.
Then one finds the relation between the beta function and the UV 
parameters at $\L_0\to \infty$
\beq\label{betaYM}
\beta(g)\equiv \frac{dg}{d \ln \mu}=
\frac{\L_0\partial_{\L_0} \br_i^\U}{\partial_g \br_i^\U}
\,.
\eeq
Similarly for the field anomalous dimensions one has
\beq
\g_i(g)=- \half[\L_0\partial_{\L_0}-\beta(g)\partial_g]\ln Z_i^\U
\,.
\eeq
Due to renormalizability the relevant parameters 
$\br_i(g,\mu/\L,\mu/\L_0)$ and $Z_i(g,\mu/\L,\mu/\L_0)$ 
are finite for $\L_0\to \infty$. Moreover they diverge for 
$\L/\mu\to \infty$. 
Therefore, as in the scalar case, in \re{betaYM} one can use the 
parameters $\br_i(g,\mu/\L,0)$ and $Z_i(g,\mu/\L,0)$ and then  
take the limit $\mu/\L\to 0$.

The one-loop expression for all the nine parameters has been
obtained in \cite{BDMYM} 
\beq\label{wf}
\eqalign{
&
\dot Z^{(1)}_A|_{\mu=0}= \frac{20}{3}\;r
\,,\;\;\;\;\;\;
\dot Z^{(1)}_c|_{\mu=0}= 2\;r
\,,\;\;\;\;\;\;
r \equiv \frac{1}{16\pi^2}
\,,
\cr&
     \dr_1^{(1)}|_{\mu=0}
\;=\;\dr_2^{(1)}|_{\mu=0}
\;=\;\dr_6^{(1)}|_{\mu=0} = 0
\,,
\cr& 
\dr_3^{(1)}|_{\mu=0} = \frac{8}{3}\;r
\,,\;\;\;
\dr_4^{(1)}|_{\mu=0} =- 2\;r
\,,\;\;\;
\dr_5^{(1)}|_{\mu=0} = -\frac{4}{3}\;r
\,.
}
\eeq
According to \re{betaYM} one finds the well known one-loop beta function 
($\beta(g)=b_0g^3+\cdots$) in the following three ways
\beq\label{beta1YM}
b_0
\;=\; (\dr_3^{(1)}-\frac{3}{2}\dot Z_A)^{(1)}|_{\mu=0}
\;=\; (\dr_4^{(1)}-\half \dot Z_A^{(1)} -\dot Z_c^{(1)})|_{\mu=0}
\;=\;\half  (\dr_5^{(1)}-2\dot Z_A^{(1)})|_{\mu=0}
\;=\;-\frac{22}{3}\;r
\,.
\eeq
The one-loop beta function cannot be obtained from 
$\dr_1^{(1)}$, $\dr_2^{(1)}$ and $\dr_6^{(1)}$ 
since they vanish to this order.

\subsection{ Flowing couplings and resummation}

We extend to the Yang-Mills case the improved perturbation discussed 
in the scalar case. 
Here we limit our discussion to the one-loop improved 
expansion. To this end we have to give the effective action to zero
loop (all fields with frequencies between $\L$ and $\L_0$)
\beq\label{zeroYM}
\eqalign{
&
\G^{(0)}[\phi,\g;\L]
=\int d^4x \biggl\{
\half \;A_\mu \cdot \partial^2 A_\mu
\;+\; W_\mu\cdot\partial_\mu c
\cr&
\;+\;\tg_{3A}(\L)\;(\partial_\nu A_\mu) \cdot A_\mu \wedge A_\nu
\;+\;\tg_{WcA}(\L)\;W_\mu \cdot A_\mu \wedge c
\;+\;\frac{\tg^2_{4A}(\L)}{4}\; (A_\mu\wedge A_\nu)\cdot
(A_\nu\wedge A_\mu)
\,,
}
\eeq
where we have introduced the three flowing couplings
$$
\tg_{3A}(\L)=\br_3(\L)
\,,\;\;\;\;\;\;
\tg_{WcA}(\L)=\br_4(\L)
\,,\;\;\;\;\;\;
\tg^2_{4A}(\L)=\br_5(\L)
\,,
$$
which are associated to the three field monomials present in the
interaction terms of the classical Lagrangian. At one loop no other
contribution will be generated (see \re{wf}).
At the physical point ($\L=0$ and $\L_0\to\infty$) one has
$$
\tg_{3A}(0)=\tg_{WcA}(0)=\tg_{4A}(0)=g
\,.
$$
By iterating the \lRG equation once, for large $\L$ one obtains 
\beq\label{zetaYM}
\eqalign{
&
\frac{\dot Z_A}{Z_A}
=\frac{19}{3}\;r\;\tg_{3A}^2
+\frac{1}{3}\;r\;\tg_{WcA}^2\,,
\cr&
\frac{\dot Z_c}{Z_c}
=\;2r\;\tg_{WcA}^2
}
\eeq
and the one-loop generalized beta function
\beq\label{GbetaYM}
\eqalign{
&
\L\partial_\L \tg_{3A}=-\frac 32 \;\frac{\dot Z_A}{Z_A}\;\tg_{3A}
-\frac{13r}{2}\tg_{3A}^3 +\frac{r}{6}\tg_{WcA}^3
+9\,r\tg_{3A}\tg_{4A}\,,
\cr&
\L\partial_\L \tg_{WcA}=-\left(\half \frac{\dot Z_A}{Z_A}
+\frac{\dot Z_c}{Z_c}\right)\tg_{WcA}
-\frac r 2\tg_{WcA}^3 -\frac{3r}2\tg_{WcA}^2\tg_{3A}\,,
\cr&
\L\partial_\L  \tg^2_{4A}=-2\;\frac{\dot Z_A}{Z_A}\;\tg^2_{4A}
+\frac{5r}{3}\tg_{3A}^4 -13\,r\tg^2_{4A}\tg_{3A}^2 +10\,r\tg_{4A}^4
\,.
}
\eeq
For $\L$ finite all terms in \re{zetaYM} and \re{GbetaYM} have to be 
multiplied by factors $f_i(\mu/\L)$ which are functions similar to 
$f(\mu/\L)$ in \re{f2} given by integrals over the cutoff
functions. These functions vanish quadratically for $\L\to 0$ and they 
tend to one for $\mu/\L\to 0$, apart for power corrections. 
Therefore one finds that for large $\L$ the three flowing couplings 
tend to the one-loop running coupling
\beq
\tg_{3A}(\L)\simeq \tg_{WcA}(\L) \simeq  \tg_{4A}(\L) \simeq g(\L)
\,,
\eeq
with
\beq\label{gYM}
g^2(k)=\frac{g^2}{1-b_0\,g^2\,\ln(k^2/\mu^2) }
\,\;\;\;\;\;\; b_0=-\frac{22}{3} r
\,.
\eeq
\begin{figure}
    \begin{center}
      \mbox{\begin{turn}{-90}\psfig{file=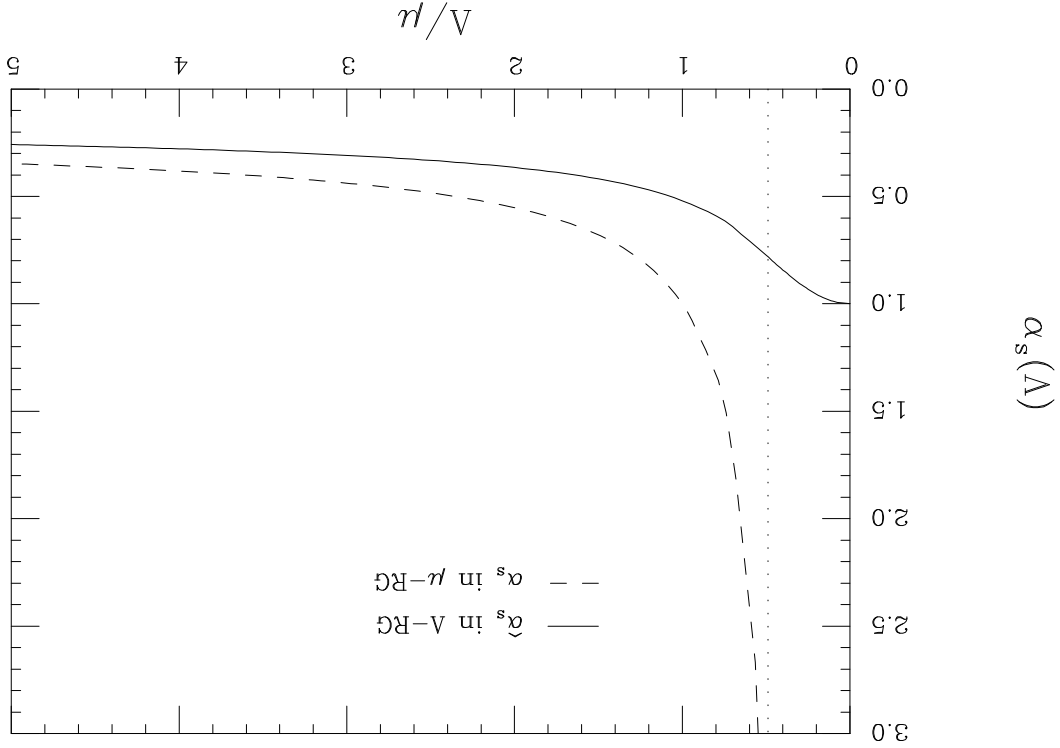,height=90 ex}\end{turn}}
    \end{center}
    \label{4}
    \caption{{\small
Plot of the one-loop running coupling $\as(\L)$ and the general pattern 
of the flowing coupling $\hat\as(\L)$ in Yang-Mills theory as function 
of $\L/\mu$. 
We use $\as(\L)\equiv g^2(\L)/4\pi$ and $\hat \as(\L)\equiv \tg^2(\L)/4\pi$
with $\as(\mu)=\hat\as(0)=1$.}}
\end{figure}
As an illustration we consider the case in which all functions 
$f_i(\mu/\L)$ to be included in \re{zetaYM} and \re{GbetaYM} 
are the same and given by the one obtained in the scalar case \re{f2}.
Then all three coupling are equal 
($\tg_{3A}(\L)=\tg_{WcA}(\L)=\tg_{4A}(\L) = \tg(\L) $)
and satisfy the RG equation 
\beq
\L\partial_\L \tg(\L)= b_0 \tg^3(\L)\;f(\mu/\L)
\,,
\;\;\;\;\;\;
\tg(0)=g
\,.
\eeq
In fig.~4 we plot the solution of this equation together with the 
corresponding running coupling $g^2(\L)$ in \re{gYM}. 
For large $\L$ we have $\tg(\L)\simeq g(\L)$. 
By decreasing the scale the power corrections in $f(\mu/\L)$ start to 
become important and the two couplings start to differ. 
The one-loop running coupling has a Landau ghost at
$k_L^2=\mu^2\,e^{1/b_0g^2}<\mu^2$.
The flowing coupling $\tg(\L)$ instead remains finite in all
range of $\L$ and reaches the physical value $g=g(\mu)$ at $\L=0$.
Thus in this case, taking $\L=0$, we can compute the physical 
vertices without problems in the IR integration region.

\section{Conclusion}
We have studied the relation between \mRG and \lRG, the formulations 
of renormalization group based on the two different ways of 
introducing the momentum scale. While in \mRG the fundamental quantity 
is the running coupling $g(k)$ (fixed to be $g$ at $k^2=\mu^2$), in 
\lRG the natural couplings are the Wilsonian flowing couplings $\tg_i(\L)$ 
(fixed to be $g$ at $\L=0$).

At large scale $\L=k$ the two couplings differ only by power 
corrections of $\mu/\L$. This implies that the characteristic 
quantities of \mRG formulation are obtained in the \lRG in the UV 
region of $\L$. In particular the beta function is given by \re{betagamma} 
and \re{betaYM} 
and is obtained by using the fact that $\tg(\L)$ does not depend on 
$\mu$ for $\mu/\L\to 0$. 
Moreover the Callan-Symanzik equation, \ie independence of $\mu$ of 
the physical quantities, corresponds to the renormalizability 
condition, \ie independence of the UV cutoff $\L_0$ of physical 
quantities for \UV. 

An interesting aspect of \lRG formulation is seen in the study of the 
problem of IR renormalons in non-Abelian gauge theory. We have shown 
that it is possible to set up an iterative procedure to solve the \lRG 
equation in such a way that one generates Feynman diagram contributions 
in which the loop momenta are ordered  ($\L^2<k_1^2\cdots<k_\ell^2$) and 
the flowing couplings at these momenta are involved.
As long as $\L$ is large, one can approximate the flowing couplings in 
terms of the running coupling. This justifies the usual assumption that 
in Feynman diagrams with a virtual momentum $k$ higher order 
corrections generate the running coupling at that scale.

As $\L$ decreases and reaches the physical point $\L=0$ one 
has to take into account that the flowing couplings differ from the 
running coupling at low frequencies. 
At low frequencies, where the perturbative running 
coupling has a Landau ghost, the flowing couplings are finite all the 
way down to zero frequence.
This allows then to perform the loop integration without problem in 
the IR region. One should observe that the fact that the flowing 
couplings are finite in the IR region is due to the presence of power 
corrections $\mu^2/k^2$ which are typical non-perturbative 
contributions.

\vspace{3mm}\noindent{\large\bf Acknowledgements}

We would like to thank C.\ Destri for helpful discussions.

\section{Appendix}
We recall the \lRG equations for the vertex functions
in the $\phi^4$ massless scalar theory in the form used in the text.
The cutoff vertex functions $\G_{2n}(p_1\cdots p_{2n};\L,\L_0)$ have
internal propagators
\beq\label{D}
D_{\L\L_0}(q)=\frac 1{q^2}\;K_{\L\L_0}(q)
\,,
\eeq
with $K_{\L\L_0}(q)=1$ for $\L^2 \le q^2 \le \L_0$ and rapidly
vanishing outside this region.

From the definition of the cutoff vertices and the fact that the
$\L$-dependence is only inside the cutoff propagator \re{D} one
deduces \cite{BDM}-\cite{Mo} the following \lRG equations (see fig.~1)
(we omit to write the $\L_0$ dependence in the vertex functions)
\beq\label{G2}
\dot \G_2(p;\L)
=\dot D^{-1}_{\L\L_0}(p)
-\half \int_q\;\G_4(q,p,-p,-q;\L)
\;\frac{1}{\G_2^2(q;\L)}
\;\dot D^{-1}_{\L\L_0}(q)
\,,
\eeq
and
\beq
\dot \G_{2n}(p_1\cdots p_{2n};\L)
=-\half \int_q\;\bG_{2n+2}(q,p_1\cdots p_{2n},-q;\L)
\;\frac{1}{\G_2^2(q;\L)}
\;\dot D^{-1}_{\L\L_0}(q)
\,,
\eeq
where the associated vertices $\bG_{2n+2}$ are
one-particle reducible with respect to combinations of momenta
including either $q$ or $-q$ and
one-particle irreducible with respect to all other combinations of
momenta.
For instance for $n=3$ one has (we omit the $\L$ and $\L_0$ 
dependence)
\beq
\bG_6(q,p_1\cdots p_4,-q)
=
\G_6(q,p_1\cdots p_4,-q)
\;-\;
\sum \G_4(q,p_1,p_2,Q_{12})
\;\frac{1}{\G_2(Q_{12})}
\G_4(-Q_{12},p_3,p_4,-q)
\eeq
with $Q_{ij}=q+p_i+p_j$ and the sum is over the six
permutations of $p_1,p_2,p_3,p_4$.
A simpler form for the \lRG equations is obtained by introducing the
kernel $I_{2n+2}(q,p_1\cdots p_{2n},-q)$ which is one-particle
irreducible and also two-particle irreducible with respect to
the two momenta $q$ and $-q$.
We have then 
\beq
\G_{2n+2}(q\cdots -q)
=I_{2n+2}(q\cdots -q)
-\half \int_{q'}
\bar I_{2n+2}(q'\cdots -q')
\;\frac{1}{\G_2^2(q';\L)}
\;\G_{4}(q,q',-q',-q)
\,,
\eeq
where, for instance for $n=3$,
\beq\label{I6}
\bar I_6(q,p_1\cdots p_4,-q)
=
I_6(q,p_1\cdots p_4,-q)
-\sum \G_4(q,p_1,p_2,Q_{12})
\;\frac{1}{\G_2(Q_{12})}
\G_4(-Q_{12},p_3,p_4,-q)
\,,
\eeq
with the sum over the permutations.

From this equation and by using \re{G2} we obtain
\beq\label{G2'}
\dot \G_2(p;\L)
=\dot D^{-1}_{\L\L_0}(p)
+\half \int_q\, I_4(q,p,-p,-q;\L)
\;\L\partial_\L\;\frac{1}{\G_2(q;\L)}
\,,
\eeq
and
\beq
\dot \G_{2n}(p_1\cdots p_{2n};\L)
=
\half \int_q\;\bar I_{2n+2}(q,p_1\cdots p_{2n},-q;\L)
\;\L\partial_\L\;\frac{1}{\G_2(q;\L)}
\,.
\eeq
In the text we discuss in particular
the equations
for the relevant parameters $\r_1(\L)$, $\r_2(\L)$ and $Z(\L)$
\beq\label{r1Z}
\eqalign{
&
\L\partial_\L (\L^2 \r_1)=\half \int_q\;
I_4(q,0,0,-q;\L)
\;\L\partial_\L\;\frac{1}{\G_2(q;\L)}
\,,
\cr&
\dot Z=\half \int_q\;
\partial_{\bp^2}I_4(q,\bp,-\bp,-q;\L)
\;\L\partial_\L\;\frac{1}{\G_2(q;\L)}
\,,
}
\eeq
and
\beq\label{r2}
\dot \r_2
=\half \int_q\;\bar I_{6}(q,\bp_1\cdots \bp_4,-q;\L)
\;\L\partial_\L\;\frac{1}{\G_2(q;\L)}
\,,
\eeq
with $\bp_i$ the subtraction point at the scale $\mu$.

The usual loop expansion is obtained by solving iteratively these 
equations with the initial condition that at zero loop the vertices 
are given by the two-point function $\G^{(0)}_2(p;\L)=D^{-1}_{\L\L_0}(p)$ 
and the four point function $\G^{(0)}_4(p_1 \cdots p_4;\L)=g$. 

\eject

\end{document}

\eject
\newpage

\begin{figcap}

\item 
Graphical representation of the \lRG equations for the two point (a) 
and $2n$-point (b) vertex functions.
The circles represent vertex functions with the IR cutoff $\L$.
The crosses represent the derivative with respect to $\L$ of the free 
cutoff propagators and vertices. 
Integration over $q$ in the loop is understood.

\item 
Feynman diagram contributing to $\dZ(\L)$ to two loop order.

\item
a) Graphical representation of the contribution to $\dr_2$ to two loop order
coming from the second term in $\bar I_6$ in \re{I6}. 
The various contributions are obtained by considering one of 
the internal vertices at one loop and the remaining at zero loop order.

b) Feynman diagram contributing to $\dr_2$ to two loop order.
This contribution is due to the first term in $\bar I_6$ in \re{I6}. 

\item 
Plot of the one-loop running coupling $g(\L)$ and the general pattern 
of the flowing couplings $\tg_i(\L)$ in Yang-Mills theory as function 
of $\L/\mu$. 
We use $\as(\L)\equiv g^2(\L)/4\pi$ and $\hat \as(\L)\equiv \tg^2(\L)/4\pi$
with $\as(\mu)=\hat\as(0)=1$.

\end{figcap}

\end{document}